%% file: Letter80211ad_v2_minor_revision.tex
\documentclass[journal,letterpaper]{IEEEtran}
\makeatletter
\usepackage{multicol}
\usepackage{cite}
\usepackage[dvips]{graphicx}
\usepackage{epsfig}
\graphicspath{{LetterEPS/}}
\usepackage{subfigure}
\usepackage{float} 
\usepackage{amsmath}
\newcommand\numberthis{\addtocounter{equation}{1}\tag{\theequation}}
\newcommand{\Nbi}{N^{\mbox\tiny{BI}}}
\newcommand{\Ncbap}{N^{\mbox\tiny{CBAP}}}
\newcommand{\Ncbapk}{N^{\mbox\tiny{CBAP}}_{\mbox\tiny{k}}}
\usepackage{array}
\usepackage{algorithm}
\usepackage{algorithmic}
\usepackage{mdwmath}
\usepackage{mdwtab}
\usepackage{psfrag}

\allowdisplaybreaks
\usepackage{url}
\newcounter{MYtempeqncnt}
\begin{document}
\title{Performance Analysis of IEEE 802.11ad MAC Protocol}
\author {\IEEEauthorblockN{Kishor Chandra, R. Venkatesha Prasad and Ignas Niemegeers}
\thanks{Authors are with Delft University of Technology, Delft, Netherlands}}
\maketitle
\begin{abstract}
IEEE 802.11ad specifies a hybrid medium access control (MAC) protocol consisting of contention as well as non-contention based channel access mechanisms. Further, it has also provisioned the use of directional antennas to compensate for the high free-space path loss at 60\,GHz.  Therefore, it significantly differs from IEEE 802.11b/g/n/ac MAC protocols and requires new methods to analyze its performance.  In this paper, we propose a new analytical model for IEEE 802.11ad MAC protocol and derive important performance metrics. The proposed model employing a three dimensional Markov chain considers all the features of IEEE 802.11ad medium access mechanism including the non-contention mode of channel access and different number of sectors due to the use of directional antennas. Analyzing IEEE 802.11ad using our model, we show that the number of sectors has a high impact on the network throughput. We also show that the MAC packet delay is significantly affected by the duration of the contention period. Our results indicate that a suitable choice of the number of sectors and the contention period can illustriously improve the channel utilization and MAC delay performance.
%
\end{abstract}
\IEEEpeerreviewmaketitle
\vspace{-2mm}
\section{Introduction}
Due to the availability of large bandwidth, millimeter wave (mmWave) frequency band (30\,GHz to 300\,GHz) has become a key enabler for the multi-Gb/s connectivity envisaged under 5G. The high free-space path loss and limited ability to diffract around obstacles require high-gain and steerable directional antennas at mmWave frequencies, which significantly impacts the design of medium access control (MAC) mechanisms~\cite{shokri2015millimeter}. The IEEE 802.15.3c for Wireless Personal Area Networks (WPANs)~\cite{iee:IEEE802.15.3c} and IEEE 802.11ad for Wireless Local Area Networks (WLANs)~\cite{IEEE802.11ad} have proposed hybrid MAC protocols operating in 60\,GHz bands. These hybrid MAC protocols consist of carrier sense multiple access with collision avoidance (CSMA/CA) and time division multiple access (TDMA) for channel access. Since the IEEE 802.11ad provides backward compatibility with the popular IEEE 802.11b/g/n/ac protocols, it has emerged as the preferred choice over the IEEE 802.15.3c. \newline
%
Compared with the IEEE 802.11b/g/n/ac Distributed Coordination Function (DCF), the IEEE 802.11ad DCF has significantly distinct features attributed to the use of directional antennas and a hybrid access mechanism.  Firstly, all the wireless stations (STAs) cannot simultaneously listen to- and hear from--the Access point (AP) due to the directional communications. Hence the area around an AP is divided into several sectors, and STAs in a sector can compete for the channel only during the allocated time period for that particular sector. Secondly, the CSMA/CA operation in a sector is suspended when either the TDMA based channel access is instantiated or when AP is busy in other sectors. Lastly, when CSMA/CA operation is suspended, backoff counter of all the involved STAs are frozen and in the next round, STAs resume the backoff process with the frozen values of backoff counters. Owing to these important differences, a thorough modelling framework of the IEEE 802.11ad MAC is needed that can take the above aspects into account.\newline 

The seminal work by Bianchi~\cite{iee:ref5} on the modelling of the IEEE 802.11 DCF employing Markov chains has been widely used for the modelling of the CSMA/CA based MAC protocols. There are several modified versions of Bianchi's model considering  various factors such as finite retransmission limit\cite{iee:ref8}, busy channel conditions\cite{Bianchi_modified_busy} and differentiated quality-of-service~\cite{Bianchi_edca}, etc. However, these models are not directly applicable in case of the IEEE 802.11ad MAC due to the provision for hybrid medium access and the use of directional antennas. There are a few studies on the performance evaluation of the IEEE 802.11ad based WLANs~\cite{PHY80211adPerformance,OFDM80211ad,11adcoperative,Hemanth80211ad}. Physical (PHY) layer performance analysis considering different modulation and coding schemes (MCS)~\cite{PHY80211adPerformance} and impact of hardware impediments\cite{OFDM80211ad} are presented without considering the impact of channel access scheme. To the best of our knowledge, only \cite{11adcoperative} and \cite{Hemanth80211ad} have attempted to build a modelling framework for the IEEE 802.11ad MAC protocol. However, the presence of non-contention channel access is not taken into account in \cite{11adcoperative}. Although, Hemanth et. al, \cite{Hemanth80211ad} have considered the non-contention part of the IEEE 802.ad MAC protocols,  the interpretation of the IEEE 802.11ad DCF is incomplete. It is assumed that after the end of every contention period, STAs refresh their backoff counters when the next contention period starts. Contrarily, this is not the case with the IEEE 802.11ad protocol, since STAs resume their backoff counters across multiple CSMA/CA periods. We propose a three-dimensional Markov chain to model the IEEE 802.11ad MAC protocol which allows us to investigate the impact of different factors attributed to the use of directional antennas and the hybrid access mechanism. Our main contributions are:  
(i)~derivation of accurate analytical expression for the channel utilization in time shared CSMA/CA; (ii)~derivation of average MAC packet delay; (iii) understanding the effects of number of sectors on channel utilization; (iv) finding the impact of contention period on the average MAC delay.
\section{IEEE 802.11\textnormal{ad} System Model}\label{intro}
\begin{figure}[]
\vspace{-5mm}
\centering
\includegraphics[width=1.1in,height=1.1in]{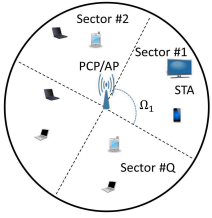} 
 \caption{IEEE 802.11ad system model.}
 \vspace{-3mm}
\label{fig_system_model}
\end{figure}

\begin{figure}[!t]
\centering
\includegraphics[width=0.3800\textwidth]{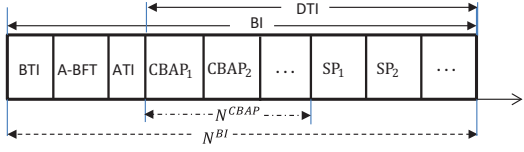}
 \caption{IEEE 802.11ad BI structure.}
 \label{fig_SF}
 \vspace{-7mm}
\end{figure}

Fig.~\ref{fig_system_model} shows an IEEE 802.11ad Personal Basic Service Set (PBSS) formed by the 60\,GHz STAs where one of them acts as PBSS Control Point/Access Point (PCP/AP). 
Fig.~\ref{fig_SF} depicts slots in an IEEE 802.11ad  beacon interval (BI) consisting of: (i)~beacon transmission interval  (BTI); (ii)~association beamforming training (A-BFT); (iii)~announcement time interval (ATI) used to exchange the management information; and (iv)~data transfer interval (DTI), consisting of contention-based access periods (CBAPs) and service periods (SPs) employing CSMA/CA and TDMA, respectively, for the channel access.\newline
%
During the CBAP, STAs in each sector access the medium in a round-robin fashion. STAs use RTS-CTS mechanism and listen in the quasi-omni (QO) mode employing wide beamwidths. The RTS and CTS frames are transmitted using QO beamwidths, hence the hidden terminal problem can be neglected. Let $m$ be the maximum retry limit and $W_{0}$  the minimum contention window size, then the window size at the $i^{th}$ retransmission stage, considering the random binary exponential backoff stage, is $W_{i}$=$2^{i}W_{0}-1$. After the  maximum transmission attempts $m$, the packet is dropped. The detailed description of the transmission procedure can be found in~\cite{IEEE802.11ad}.\newline
Let us assume that $n$ STAs are deployed in $Q$ sectors around the PCP/AP. We assume uniformly distributed STAs to keep our analysis simple. Let the $k^{th}$ sector having a beamwidth of $\Omega_k$ have $n_{k}$ STAs where $\sum_{k=1}^{k=Q}\Omega_k=2\pi$ and  $\sum_{k=1}^{k=Q}n_k=n$. Let $\tau_k$ and $p_k$ be the transmission and collision probabilities of a packet in the $k^{th}$ sector, respectively. $p_k$ is also known as the \textit{conditional collision probability} under steady state which is independent of the number of retransmission attempts in a sector. For brevity, henceforth, we will represent $\tau_k$ and $p_k$ by $\tau$ and $p$, respectively; though each sector can have different $\tau$ and $p$ if they have a different number of STAs. Using the above definitions under saturation condition, we have,
\small
\begin{equation}\label{eq:collision}
p=1-\left(1-\tau\right)^{n_{k}-1}.
\end{equation}\normalsize
\begin{figure}[!t]
\vspace{-5mm}
\includegraphics[width=0.4800\textwidth]{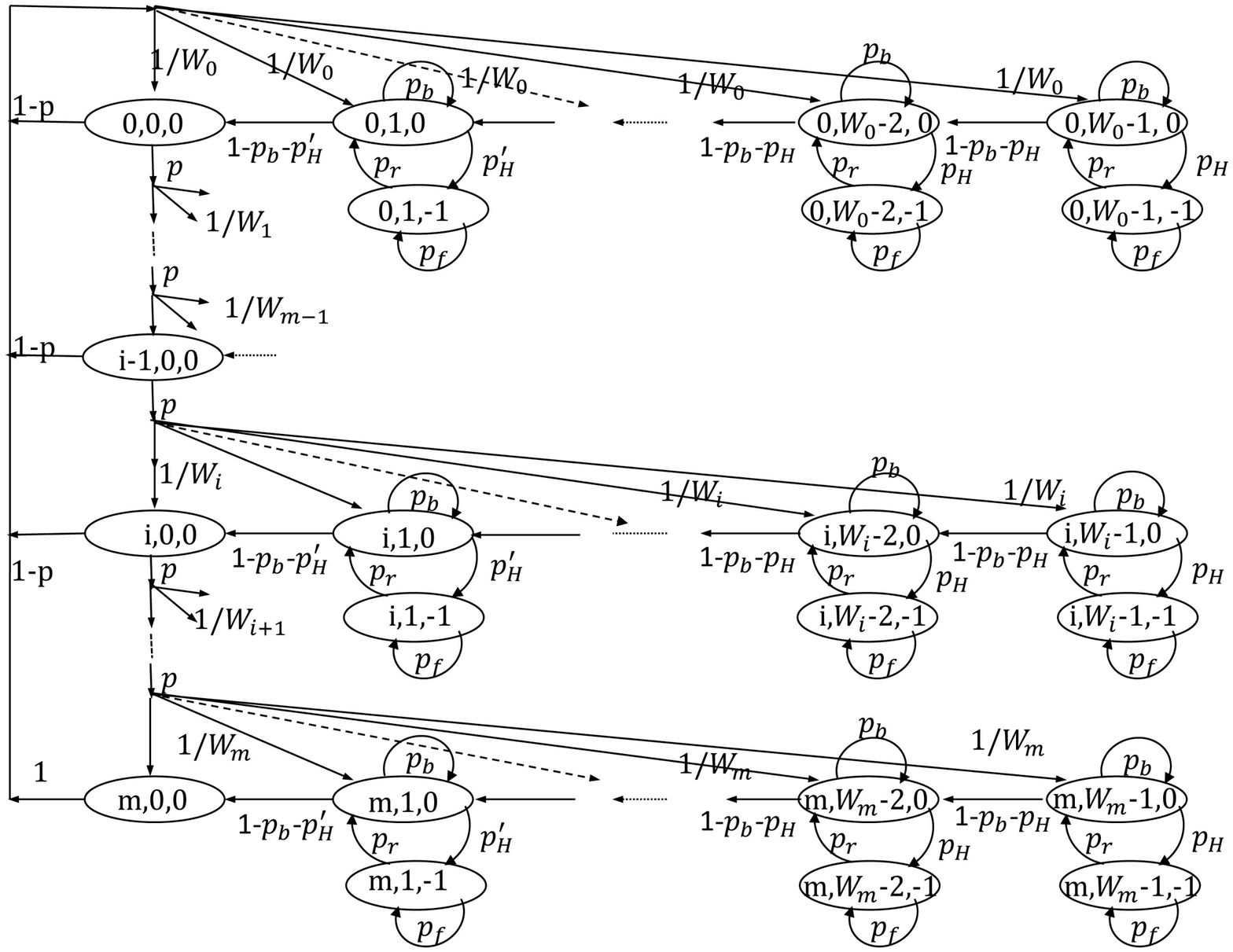}
 \caption{Markov chain model for packet transmission states.}
\label{fig_markov}
\vspace{-5mm}
\end{figure}
\vspace{-6mm}
\section{Three Dimensional Markov Chain Model for Packet Transmission}
We define a three dimensional Markov chain as shown in Fig.~\ref{fig_markov} where each state is represented by the triplet $(s(t),b(t),h(t))$. Here, $s(t)$ represents the backoff stage $i$, $i\in[0,m]$ and $b(t)$ represents the residual backoff time counter $j$, $j\in[0,W_i-1]$. To differentiate between contention and non-contention part of BI, we define $h(t)$, 
\small
\begin{equation}\label{trasarray0111}
  h(t)=\begin{cases}
  \begin{array}{ll}
  0,&\mbox{if packet is part of an ongoing CBAP}\\
 -1,&\mbox{otherwise }
  \end{array}
  \end{cases}
  \end{equation}
  \normalsize
  \input{EquationB00inFigure.tex}
  \vspace{-5mm}
\subsection{Transition probabilities} 
Let \small$\Nbi$\normalsize and \small$\Ncbap$\normalsize be the lengths of BI and CBAP, respectively. Let the $k^{th}$ sector have a CBAP length of $\Ncbapk$ \small$(\sum_{k=1}^{k=Q}\Ncbapk=\Ncbap)$\normalsize. Let $N^{\mbox{\tiny{F}}}$ be the time required to successfully transmit a data frame. The time durations used here are measured in unit of slot time $\sigma$. According to the IEEE 802.11ad MAC protocol, when $j\in[2, W_i-1]$, STAs jump to non-contention state if CBAP time counter reaches zero. However, if $j=1$,  STAs have to avert the packet transmission and jump to non-contention state even if allocated CBAP duration is not finished yet but the CBAP counter has reached a value lower than the total required time to transmit a packet. Hence, the transition probabilities of going from contention to non-contention state are given by,
\small
\begin{equation}\label{trasarray0}
   P\{i,j,-1|i,j,0\}=\begin{cases}
  \begin{array}{ll}
  p_H;&j\in[2, W_i-1], i\in[0,m]  \\
 p'_H;&j=1, i\in[0,m]\\
  \end{array}
  \end{cases}
  \end{equation}
  \normalsize
For the STAs in the $k^{th}$ sector we have $p_H=\frac{1}{\Ncbapk}$ and $p'_H=\frac{N^{\mbox{\tiny{F}}}}{\Ncbapk}$. Let $p_r$ be the probability of transition from  a non-contention state to a contention state and $p_f$ be the transition probability of staying in a non-contention state, then,
 \small
  \begin{subequations}  
  \begin{align}
  P\{i,j,0|i,j,-1\}=p_r;\ \ \ &j\in[1, W_i-1], i\in[0,m], \label{trasarray61} \\
  P\{i,j,-1|i,j,-1\}=p_f;\ \ \ &j\in[1, W_i-1], i\in[0,m], \label{trasarray62} 
  \end{align}
  \end{subequations}
  \normalsize
   where, $p_r=\frac{\Ncbapk}{\Nbi}$ and $p_f=1-p_r$.
   Let $p_b$ be the state transition probability of channel being busy during the contention period, then,
 \small
  \begin{align}\label{trasarray63}
  P\{i,j,0|i,j,0\}=p_b;\ \ &j\in[1, W_i-1], \ \ i\in[0,m].
  \end{align} 
  \normalsize
  An STA observes  busy channel when at least one STA among the remaining $n_k-1$ STAs occupies the channel. Hence  $p_b$, i.e., probability of not decreasing the backoff counter during CBAP period, is calculated as, \small$p_b=1-(1-\tau)^{n_k-1}$\normalsize.  The state transitions probabilities of a successful transmission and collision are, 
   \small
  \begin{subequations}
  \begin{align}
   P\{0,j,0|i,0,0\}=(1-p)/{W_0}; j\in[0,W_i-1], i\in[0,m].\label{trasarray71}\\  
   P\{i,j,0|i-1,0,0\}=p/W_i; j\in[0, W_i-1], i\in[0,m]. \label{trasarray72}\\\nonumber  
  \end{align}
  \end{subequations} 
  \normalsize
  \vspace{-2pt}
The state transition probability due to decrementing the backoff counter is given by,
   \small
   \begin{equation}\label{trasarray05}
  P\{i,j-1,0|i,j,0\}=\begin{cases} 
  \begin{array}{ll}
  1-p_b-p_H; &j\in[2,W_i-1], i\in[0,m]  \\
  1-p_b- p'_H; &j=1, i\in[0,m]\\
  \end{array}\nonumber
    \end{cases}
  \end{equation}
  \normalsize
  \vspace{-5mm} 
 \subsection{Steady state probabilities}
For $i\in[1,m]$, from  Fig.\ref{fig_markov} we obtain,
\small
 \begin{subequations} \label{EquationBIJ} 
  \begin{align}
  b_{i,j,0}&=\frac{p(W_i-j)b_{i-1,0,0}}{(1-p_b-p_H)W_i}; \ \ \ i\in[1,m], j\in[2,W_i-1]. \label{Equation_backto_b_000_5_shortVersion} \\
b_{i,1,0}&=\frac{p(W_i-1)}{(1-p_b-p'_H)W_i}b_{i-1,0,0};\ \ \ i\in[1,m], j=1. \label{Equation_backto_b_000_6shortVersion} 
  \end{align}
  \end{subequations}
\normalsize
%
\vspace{-8pt}
\small
\begin{equation}
\hspace{-42mm}
b_{i,0,0}=p^ib_{0,0,0}; \ \ \ i\in[1,m].
\end{equation}
\normalsize
Relationship amongst steady state probabilities for $i=0$, 
\small
 \begin{subequations}\label{LastBIJ} 
\begin{align}
b_{0,j,0}&=\frac{(1-p)({W_0-j})}{(1-p_b-p_H){W_0}}\sum_{i=0}^{m}b_{i,0,0}; \ \ \ j\in[2, W_0-1].\label{Equation_backto_b_000_9shortVersion}\\
b_{0,1,0}&=\frac{1-p}{1-p_b-p'_H}\frac{W_0-1}{W_0}\sum_{i=0}^{m}b_{i,0,0}; \ \  j=1.\label{Equation_backto_b_000_10shortVersion}
\end{align}
\end{subequations}
\normalsize
 \vspace{-1pt}
Since the sum of the steady state probabilities should be 1, hence,
\small
\begin{align}\label{totalSum_markovchain}
 \sum_{i=0}^{m}\sum_{j=0}^{W_i-1}\sum_{k=-1}^{0}b_{i,j,k}=1. 
 \end{align}
 \normalsize
 Expanding (\ref{totalSum_markovchain}) and using (\ref{EquationBIJ})\,--\,(\ref{LastBIJ}), we obtain  (\ref{new_analysis_Markovchain}), where, $\eta'=\left(1+\frac{p_H'}{1-p_f}\right)\frac{1}{1-p_b-p_H'}$ and $\eta=\left(1+\frac{p_H}{1-p_f}\right)\frac{1}{1-p_b-p_H}$.
From Fig.~\ref{fig_markov}, transmission probability $\tau$ is the sum of the steady state probabilities of being in the head-of-line states, 
\setcounter{equation}{11}
\small
\begin{align}
\tau=\sum_{i=0}^{m}b_{i,0,0}=\frac{1-p^{m+1}}{1-p}b_{0,0,0},\label{eq23}
\end{align}
\normalsize
where, $b_{0,0,0}$ is given by~(\ref{new_analysis_Markovchain}). Given this relation, (\ref{eq:collision}) and ~(\ref{eq23}) can be solved for $p$ and $\tau$. 
\vspace{-5pt}
\subsection{Channel utilization}
During the contention process, the probability of, a slot being idle, having a successful transmission or a collision, can be expressed as \small$P_{\mbox{idle}}=\left(1-\tau\right)^{n_{k}}$, $P_{\mbox{suc}}=n_{k}\tau(1-\tau)^{n_{k}-1},$\normalsize and \small$P_{\mbox{col}}=1-P_{\mbox{idle}}-P_{\mbox{suc}}$\normalsize, respectively. Let $T_{\mbox{idle}}=\sigma$, $T_{\mbox{suc}}$ and $T_{\mbox{col}}$ be the duration of an idle time slot, a successful transmission and a failed transmission, respectively. Then,  the channel utilization or the normalized throughput $U_{k}$ of the $k^{th}$ sector, which is defined as the fraction of the total allocated CBAP time (${\Ncbapk}$) the channel is used in that sector for the successful packet transmissions, is given by,
\small
\begin{align}\label{total1}
U_{k}&=\frac{P_{\mbox{suc}}E[P]}{P_{\mbox{idle}}T_{\mbox{idle}} + P_{\mbox{suc}}T_{\mbox{suc}} + P_{\mbox{col}}T_{\mbox{col}}},
\end{align}
\normalsize
where \small$E[P]$\normalsize\, is the average duration of a payload packet and $T_{\mbox{idle}}$ is duration of one slot.
Here, \small$T_{\mbox{suc}} =T_{\mbox{rts}} + 2\mbox{SIFS} + T_{\mbox{cts}}+\mbox{DIFS}+T_{\mbox{data}}+T_{\mbox{ack}}$\normalsize\, and  \small$T_{\mbox{col}} = T_{\mbox{rts}} + \mbox{SIFS} + \mbox{DIFS + RIFS}$\normalsize, where \small$T_{\mbox{rts}}$\normalsize\, and  \small$T_{\mbox{cts}}$\normalsize\, are the durations of the RTS and CTS frames, respectively.  \small$\mbox{DIFS}$\normalsize\, is the Distributed Backoff Inter-Frame Space and $\mbox{RIFS}$ is the Retransmission Inter-Frame Space. We have neglected the sector switching time (it is around 3 to 18$\mu$s), because it is negligible compared to the allocated CBAP time (which would be at least few milli seconds) for a given sector. Since the sector-wise medium access during CBAP is carried in a round-robin fashion, the average channel utilization during whole CBAP period (considering all the sectors) is given by \small$U=\frac{1}{\Ncbap}\sum_{k=1}^{k=Q}U_k\Ncbapk$.\normalsize
\subsection{MAC delay analysis}
 MAC delay $E[D]$ is the expected time between the arrival of a packet at the MAC layer  until it is successfully transmitted. The probability that a packet is discarded after $m$ backoff stages is given by $p^{m+1}$. Hence the  probability of a successful transmission in the $i^{th}$ backoff stage is expressed as \small$P(TX=i|success)=\frac{p^i(1-p)}{1-p^{m+1}}$\normalsize. Note that the first transmission attempt corresponds to the zeroth backoff stage. Let $E[D_i]$ be the average delay encountered by a packet before successfully transmitted in $i^{th}$ backoff stage, then the average MAC delay, $E[D]$, is,
\small
\begin{align}\label{equation:overallDelay_shortVersion}
E[D]=\sum_{i=0}^{m}P(TX=i|\mbox{success})E[D_i].
\end{align} 
\normalsize
Here, $E[D_i]$ consists of: (i)~the delay accumulated during $i$ collisions, (ii)~the time taken by successful transmission in $i^{th}$ backoff stage, and (iii)~the backoff process delay corresponding to $i+1$ backoffs. The backoff process delay is composed of average delay incurred when the medium is busy because of other transmissions, backoff counter decrement during idle channel conditions, and delay caused due to the STA being in non-contention period. Thus, $E[D_i]$ can be expressed as,
\small
\begin{align}\label{Equation:averageidleslot_1_shortVersion}
E[D_i]=iT_{\mbox{col}}+T_{\mbox{suc}}+\sum_{z=0}^{z=i}\frac{W_z-1}{2}\frac{\sigma_{avg}}{1-p_b-p_H},
\end{align}
\normalsize
where $\sigma_{avg}$ is defined as one-step state transition period and can be calculated as         \small$\sigma_{avg}=(1-p_H)(P^o_{\mbox{idle}}\sigma+P^o_{suc}T_{\mbox{suc}}+P^o_{\mbox{col}}T_{\mbox{col}})+p_H(\Nbi-\Ncbapk)\sigma$\normalsize. Here \small$P^o_{\mbox{idle}}=(1-\tau)^{n_k-1}$, $P^o_{\mbox{suc}}=(n_k-1)\tau(1-\tau)^{n_k-2}$\normalsize \ and \small$P^o_{\mbox{col}}=1-P^o_{\mbox{idle}}-P^o_{\mbox{suc}}$\normalsize \ represent the probability of channel being observed idle, busy with a successful transmission, and busy with a collision, respectively.
\begin{figure}[!t]
\vspace{-5mm}
\hspace{-10mm}
\subfigure[Throughput Vs $W_0$.]{ 
\psfrag{x}[cc][cc][0.80]{{Number of STAs}}
\psfrag{y}[cc][cc][0.80]{{Normalized throughput}}
\includegraphics[width=4.6cm, height=4cm]{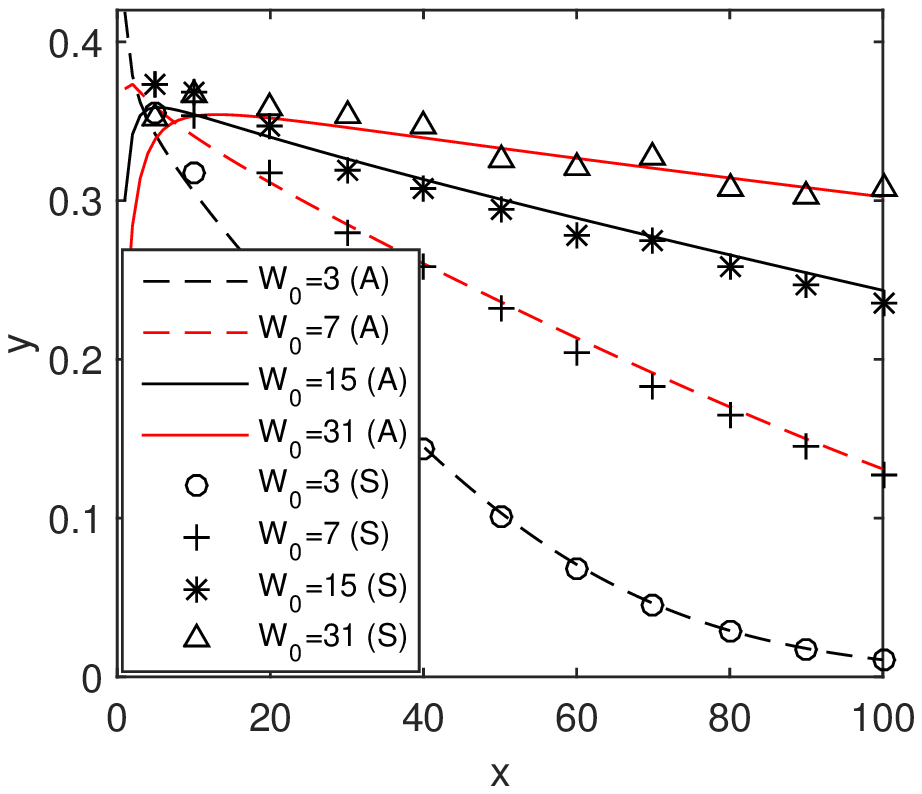} 
\label{fig_Throughput_window3_31_05_sector1}
}
\subfigure[Throughput Vs number of sectors.]{
\psfrag{x}[cc][cc][0.80]{\small{Number of STAs}}
\psfrag{y}[cc][cc][0.80]{\small{Normalized throughput}}
\includegraphics[width=4.6cm, height=4cm]{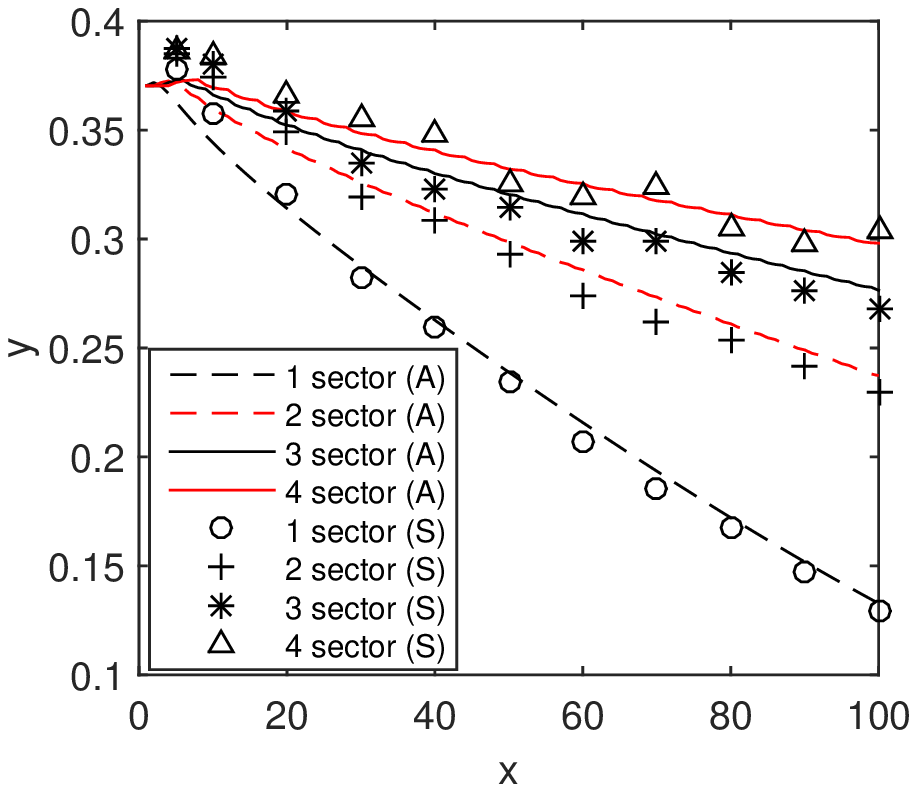}
\label{fig:Throughput_1505_50_sector1_4}
}
\subfigure[Throughput Vs CBAP duration.]{
\hspace{-12mm}
\psfrag{x}[cc][cc][0.80]{\small{Number of STAs}}
\psfrag{y}[cc][cc][0.80]{\small{Normalized throughput}}
\includegraphics[width=4.6cm, height=4cm]{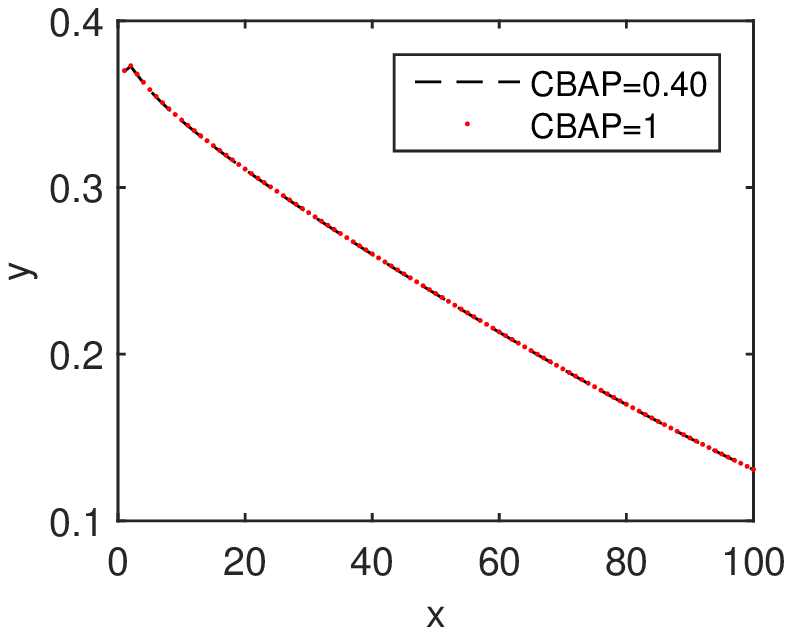}
\label{fig:Throughput_CBAPfraction1505_sector1}
}
\subfigure[Delay Vs CBAP duration.]{ 
\psfrag{x}[cc][cc][0.80]{\small{Number of STAs}}
\psfrag{y}[cc][cc][0.80]{\small{Average MAC delay(ms)}}
\includegraphics[width=4.6cm, height=4cm]{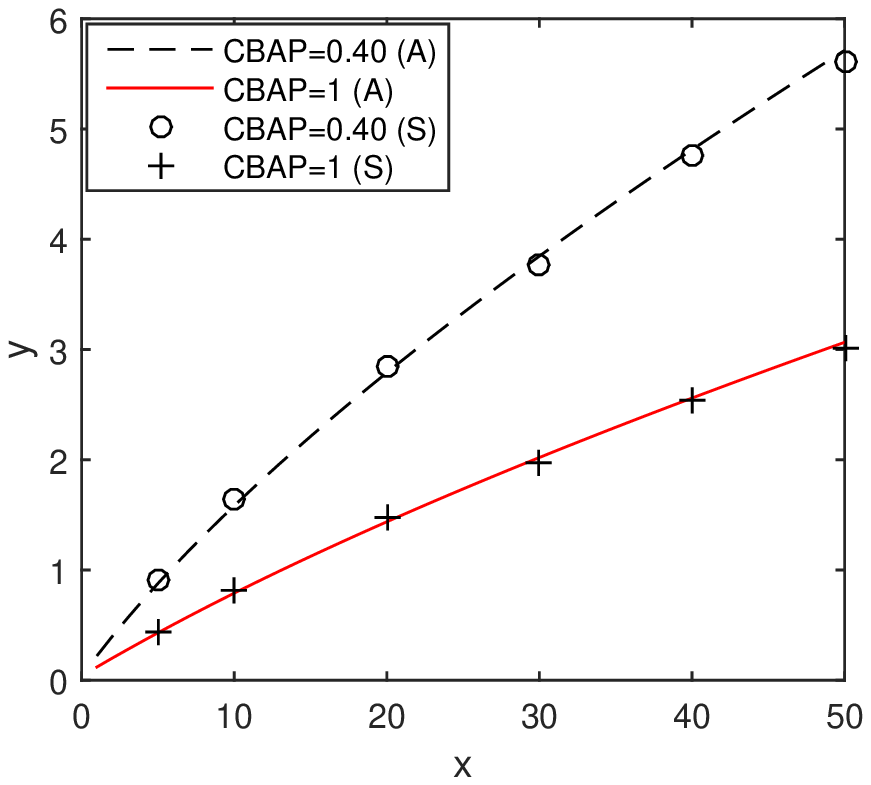}
\label{fig:delay_15_05_sector1_cbap40_100_7995bytes}
}
\caption{Throughput and delay performance.}
\label{fig:Throughput_window_sector_cbap}
\vspace{-5mm}
\end{figure} 
\vspace{-5mm}
\section{Numerical results}
Fig.~\ref{fig:Throughput_window_sector_cbap} shows the simulation (S) and analytical (A) results for the throughput and delay performance. We built a discrete event simulator in MATLAB using the IEEE 802.11ad specifications. We assumed an ideal flat-top directional antenna with main lobe gain ($G=\frac{2\pi}{\omega}$) without any side lobes. The IEEE 802.11ad Control PHY is used to transmit the RTS, CTS and ACK frames, while MCS4 with datarate of 2\,Gb/s is used to transmit data.  We considered a uniform distribution of STAs around the AP considering a radius of 10\,m.  The size of a data frame is 7995\,Bytes, which is the maximum MSDU size specified in IEEE 802.11ad. The size of BI is 100\,ms and CBAP fraction is 0.40. The packet retry limit is set to 5 and $W_0=7$. Simulations results are the average of 10000 runs in each case. As explained earlier we consider packet saturation at MAC queue of each STA which implies that every STA has a head-of-line packet ready to join contention process after the current packet is transmitted. All other parameters used in the numerical evaluation are listed in the Table~\ref{table_CAP}.

Fig.~\ref{fig_Throughput_window3_31_05_sector1} shows the normalized throughput as a function of the number of STAs for varying $W_0$ in a single sector. We can see that when the number of STAs is increased, larger $W_0$ results in a better channel utilization as the probability of collision decreases with an increase in $W_0$. This is a well-established result with respect to IEEE 802.11b/g. However, a large $W_0$ would also result in more time wasted being idle~\cite{iee:ref8}.  Fig.~\ref{fig:Throughput_1505_50_sector1_4} shows the impact of the number of sectors on the CBAP throughput with  $W_0=7$. It is evident that when the number of STAs is increased, having more sectors is beneficial. This is due to the fact that at an instant, simultaneously contending STAs would be divided by a factor equal to the number of sectors thus resulting in less collisions and hence better channel utilization. For example when the number of STAs is 40, using three or four sectors gives better throughput performance than one (omni) or two sectors.  Hence, when the number of STAs increases, instead of using a large $W_0$, having more sectors is better because: (i) only PCP/AP need to adjust its beamwidth while in the former case all the STAs need to adapt their $W_0$ and (ii) the wasted idle period using large $W_0$ can be avoided. However, when there are fewer STAs, using less number of sectors gives better throughput. Hence, intelligently determining the sector beamwidth will ensure better utilization of channel.

To assess the impact of CBAP duration, we normalized CBAP with respect to BI duration~$(\frac{\Ncbap}{\Nbi})$. We used CBAP values of 0.40 and 1. It can be seen that throughput is unaffected with respect to CBAP. This is because CBAP throughput mainly depends on $n$, $W_0$ and $m$. The amount of time when STAs are out of CBAP is not considered to calculate the throughput since during that time either STAs in other sectors would contend for the channel or STAs would be utilizing the TDMA based access periods. On the other hand, CBAP has a significant impact on the delay experienced by a packet as can be seen in Fig.~\ref{fig:delay_15_05_sector1_cbap40_100_7995bytes}. We can see that for CBAP value of 0.40, average packet delay is significantly higher compared with the CBAP value of 1. This is because in the former case, if allocated CBAP period expires before a packet is transmitted, the packet will have to wait until the next round of CBAP to access the medium. Therefore selection of the duration of contention part of BI would play an important role in determining the channel access delay
\begin{table}[!t]
\renewcommand{\arraystretch}{1.0}
\caption{CBAP Analysis Parameters.}
\label{table_CAP}
\centering
\begin{tabular}{|c|c|c|c|c|c|}
\hline
RTS & {CTS} & {ACK}&SIFS&RIFS&DIFS\\
\hline
20 Octets & 26 Octets & 14 Octets  & 2.5 $\mu$s & 9 $\mu$s & 13.5 $\mu$s\\
\hline
\end{tabular}
\vspace{-5mm}
\end{table}
\section{Conclusion}\label{remark} 
The analytical models developed for IEEE 802.11b/g/n/ac MAC protocol cannot be directly applied to the IEEE 802.11ad due to the use of directional antennas and hybrid access periods. Thus we presented a new and a thorough IEEE 802.11ad MAC protocol model using three dimensional Markov chain. We analyzed the dependencies of contention period and different number of sectors on the MAC delay and throughput. It is shown that the number of sectors has a significant impact on the channel utilization. For example, using four sectors resulted in up to 30-50\% improvement in the channel utilization  when the total number of STAs varies from 30 to 50.  We have also shown that the average MAC delay is doubled when the CBAP is reduced to 40\% from 100\% of total BI duration with 50 contending STAs. 
\tiny
\bibliographystyle{IEEEtran}
\bibliography{Adaptreferences}
\end{document}

%% file: EquationB00inFigure.tex
 \begin{figure*}[]
    \small
    \setcounter{MYtempeqncnt}{\value{equation}}
    \setcounter{equation}{10}
        \begin{equation}
{b_{0,0,0}={\bigg(1+\frac{W_0-1}{W_0}\left(\eta'+\eta\frac{W_0-2}{2}\right)(1-p^{m+1})+p\frac{1-p^m}{1-p}\left(1+\eta'-\frac{3}{2}\eta\right)+\frac{p}{2W_0}\frac{1-(\frac{p}{2})^m}{1-\frac{p}{2}}(\eta-\eta')+\eta pW_0\frac{1-(2p)^m}{1-2p}\bigg)^{-1}}\label{new_analysis_Markovchain}}
    \end{equation}
    \setcounter{equation}{\value{MYtempeqncnt}}
    \hrulefill
    \normalsize
    \end{figure*}